\documentclass[journal]{journal}
\ifCLASSINFOpdf
\else
\fi
\usepackage{graphicx}
\usepackage{amssymb}
\usepackage{amsmath,amssymb,amsfonts,amsthm}
\usepackage{mathtools}
\usepackage{graphicx}
\usepackage{tikz}

\usepackage{bbm}
\usepackage{booktabs}
\usepackage{listings}
\usepackage{amsfonts}
\usepackage{amsmath}
\usepackage{wrapfig}
\usepackage{pgfplots}
\usepackage{pgfplotstable}
\usepackage{adjustbox}
\usepackage{subcaption}
\usepackage{ifluatex}
\ifluatex
\usepackage{fontspec}
\usepackage[english]{selnolig}
\fi

\usepackage[none]{hyphenat}
\usepackage[justification=centering]{caption}	
\usepackage[labelsep=space]{caption}		
\usepackage[font=footnotesize]{caption} 
\newtheorem{definition}{Definition}
\begin{document}

\title{Process Comparison Using Object-Centric Process Cubes}

\author{Anahita~Farhang~Ghahfarokhi,
        Alessandro~Berti,
        Wil~M.P.~van~der~Aalst% <-this % stops a space

\thanks{Anahita~Farhang~Ghahfarokhi, Alessandro~Berti, Wil~M.P.~van~der~Aalst are with the Process and Data Science institute, RWTH Aachen University, Germany (e-mail: farhang@pads.rwth-aachen.de, a.berti@pads.rwth-aachen.de, wvdaalst@pads.rwth-aachen.de).}% <-this % stops a space
}

\markboth{Journal of \LaTeX\ Class Files,~Vol.~6, No.~1, January~2007}%
{Shell \MakeLowercase{\textit{et al.}}: Bare Demo of IEEEtran.cls for Journals}



\maketitle
\thispagestyle{empty}

\begin{abstract}
%\boldmath
Process mining provides ways to analyze business processes. Common process mining techniques consider the process as a whole. However, in real-life business processes, different behaviors exist that make the overall process too complex to interpret. Process comparison is a branch of process mining that isolates different behaviors of the process from each other by using process cubes. Process cubes organize event data using different dimensions. Each cell contains a set of events that can be used as an input to apply process mining techniques. Existing work on process cubes assume single case notions. However, in real processes, several case notions (e.g., order, item, package, etc.) are intertwined. Object-centric process mining is a new branch of process mining addressing multiple case notions in a process. To make a bridge between object-centric process mining and process comparison, we propose a process cube framework, which supports process cube operations such as slice and dice on object-centric event logs. To facilitate the comparison, the framework is integrated with several object-centric process discovery approaches.
\end{abstract}


% \fi
\IEEEpeerreviewmaketitle

\section{Introduction}
\IEEEPARstart{E}{very} organization has to manage business processes such as the Purchase-to-Pay (P2P) and Order-to-Cash (O2C) processes. To manage real-life processes we need to consider two challenges. First, the nature of the business processes is not static due to the environmental changes (e.g., seasonal demands, changing in customer preferences). Thus, different behaviors may exist in business processes~\cite{bolt2015multidimensional}. We can compare different process behaviors using process cubes. Process cubes are inspired by the notion of OLAP~\cite{ribeiro2011event}. Similar operations to OLAP are defined for process cubes, i.e., slice, dice, drill-down, and roll-up~\cite{chen2009graph}. 

Second, multiple objects interact with each other in the business processes~\cite{van2019object}. Considering simple P2P and O2C processes, multiple objects, e.g., order, item, and package are involved. Moreover, different relations exist between these objects. For example, an order contains multiple items and multiple items are packed in one package for the delivery. Each of these objects can be considered as a case notion (i.e., a process instance). Therefore, in real business processes, extracted from ERP systems, we have multiple case notions. However, like most existing process mining techniques, current process cube approaches can only handle one case notion at a time~\cite{bolt2015multidimensional,vogelgesang2013multidimensional,van2013process}. Thus, they cannot cover multiple case notions involved in business processes such as orders and items. The interaction between multiple case notions can be analyzed using object-centric process mining~\cite{van2019object}. This emerging subfield of process mining provides a more general vision of business processes by considering multiple case notions in processes.

In this paper, we address the problem of handling multiple interacting case notions in process cubes. In order to fit object-centric event logs into process cubes, some challenges arise due to the nature of object-centric event logs, where we need to consider multiple case notions. In traditional event logs, each event refers to a single case notion, an activity, a timestamp, and some possible additional attributes (e.g., location, cost, etc.). However, in object-centric event logs, each event may refer to multiple case notions. For example, suppose we have an event with a confirm order, related to one order but multiple items. In fact, there is a one-to-many relationship between order and item. Then, each event is related to one order and many items, i.e., convergence~\cite{van2019object}. Considering slicing based on item dimension, the values of item dimension are not atomic. However, in the existing process cubes~\cite{bolt2015multidimensional,vogelgesang2013multidimensional,andreswari2019olap,nik2019bipm}, we need atomic values to apply each operation.

In this paper, we extend the notion of process cubes to be applicable to non-atomic values and provide an open-source implementation of our object-centric process cube approach. Moreover, to facilitate the comparison, we use object-centric process discovery approaches. We can extract process models of the cells of the process cubes and by comparing them against each other, we are able to do performance analysis from different angles.

\begin{figure}[t]
	\centering
	\includegraphics[scale=0.37]{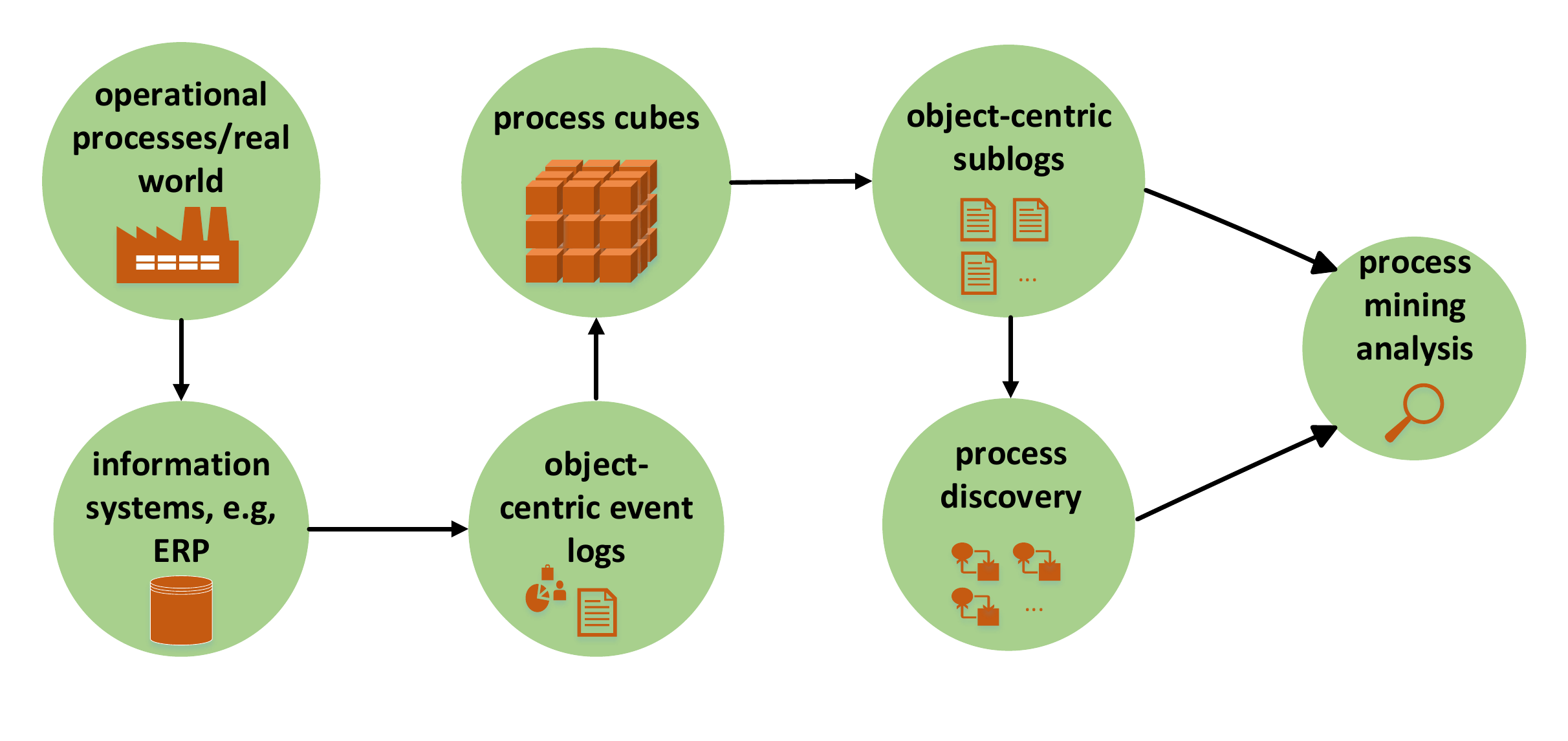} % bayad nesbate 3505 be 2569 hefz shavad=1.36
	\caption{Schematic overview of the proposed approach. In the proposed framework, we feed the process cube with object-centric event logs}
	\label{fig-merge-worlds}
	\centering
\end{figure}

A schema of the proposed approach is shown in Fig.~\ref{fig-merge-worlds}. In the first step, we have operational and business processes running in the real world. The information of these business processes is recorded in information systems such as SAP or ERP systems. We can extract object-centric event logs, i.e., event logs with multiple case notions from data stored in information systems. Then, we use the object-centric event log as an input for the process cube. In the process cube, different operations such as slicing and dicing are applicable. Through the application of different operations, we obtain object-centric sublogs. We can apply the object-centric process discovery methods on the object-centric sublogs to have the model of the process. Using such object-centric sublogs and the discovered models, we can analyze the process and find the pitfalls of the process.

The structure of the paper is as follows. In Section~\ref{Related Work}, we discuss related work. In Section~\ref{Running Example}, we describe the running example that is used throughout the paper. In Section~\ref{Process Cubes}, we extend the definitions of the process cube to support object-centric event logs. Section~\ref{Implementation} provides our implementation of the proposed framework. In Section~\ref{Evaluation}, we provide performance related results of our framework. Finally, Section~\ref{Conclusion} concludes the paper and discusses the future work.

\section{Related Work}\label{Related Work}
In this section, first, we present the work related to object-centric process mining. Then, we discuss the developed approaches on process comparison. 

One of the approaches developed to model the processes with multiple case notions is artifact-centric modeling. Artifacts combine process and data as the building blocks~\cite{cohn2009business}. In~\cite{bhattacharya2007towards}, the authors formulate artifact-based business models and develop an initial framework that enables the automated construction of the processes. In~\cite{lu2015discovering}, the authors introduce an artifact-centric process model, showing business objects and their life-cycles. The proposed techniques do not show the whole process in one diagram, which leads to losing a general vision over the process. Object-Centric Behavioral Constraint (OCBC) models show the whole process in one diagram and incorporate data perspective in the process model~\cite{li2017automatic}. The main challenge for OCBC models is complexity, which leads to the development of MVP (Multiple Viewpoint) Models~\cite{berti2020extracting}. MVP models are graphs,  annotated by frequency and performance information, where relationships between activities are shown by colored arcs. Object-centric Petri nets are another type of object-centric process models that can be extracted from object-centric event logs which provide the execution semantics~\cite{hernandez2015handling}. In this paper, comparison between the MVP models and Object-centric Petri nets extracted from cube operations against the whole cube is possible, which is helpful in process analysis.

Process cubes are inspired by OLAP, where event data are organized using dimensions. Each cell contains events, which can be used as an input to apply process mining techniques such as process discovery. In~\cite{ribeiro2011event}, the event cube is introduced with the application of OLAP operations on event data. The first notion for process cubes was proposed in~\cite{van2013process} and then enhanced in~\cite{bolt2015multidimensional}. In~\cite{bolt2015multidimensional} an approach for interactive analysis of the event data is also proposed. Process cubes were used for analysis in several case studies for different purposes~\cite{andreswari2019olap,vogelgesang2013multidimensional,gupta2014process}. Although, none of them addresses handling object-centric event logs.

\section{Running Example}\label{Running Example}
\begin{figure}[b]
	{ 
		\centering
		\includegraphics[scale=.3]{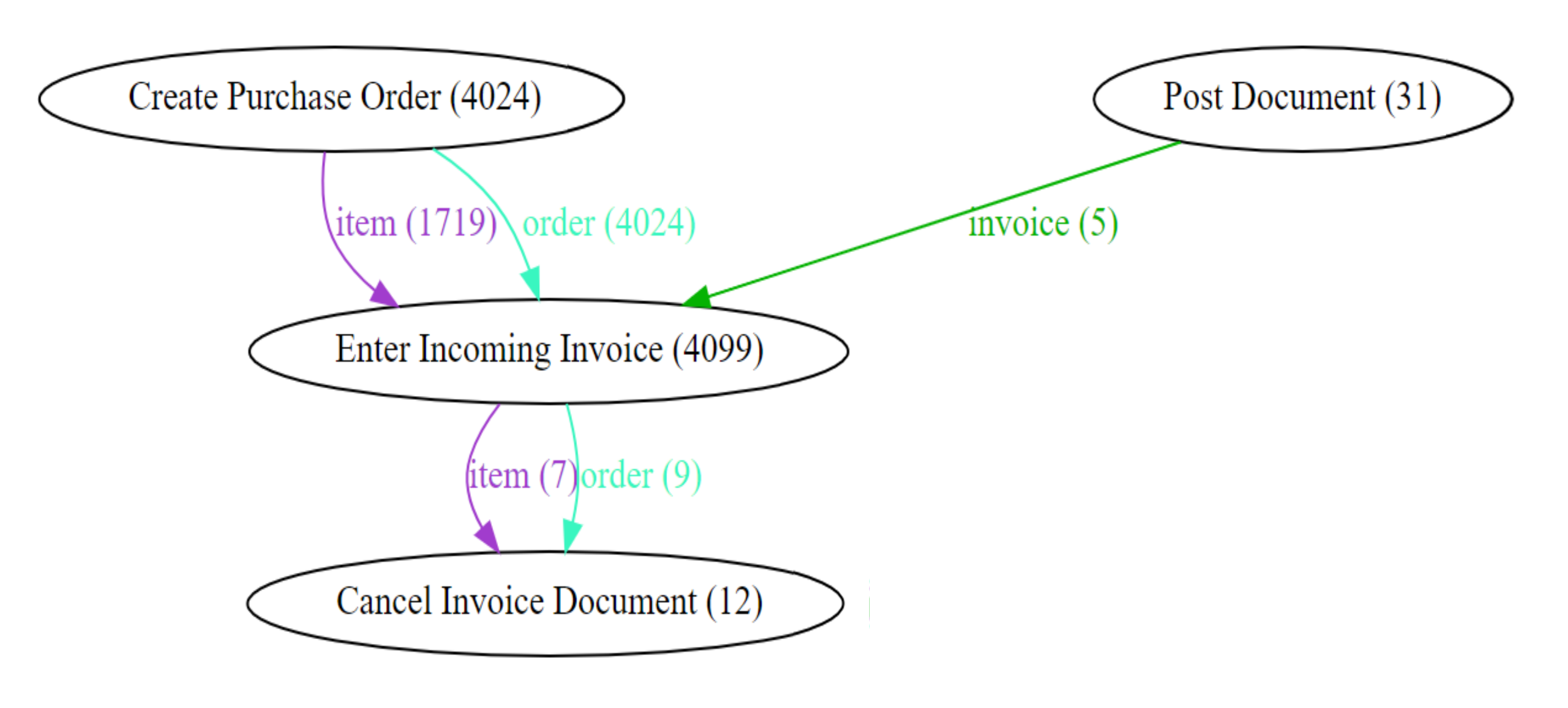}
		\caption{MVP model for a fragment of the process}
		\label{fig-obj-model}
	}
\end{figure}

The example process is from the publicly available SAP IDES system which belongs to a real purchasing process (available in https://gitlab.com/Anahita-Farhang/process-cube). It contains 17500 events, 4 object types (i.e., case notions), 4 attributes, and 10 number of activities recorded from 2000 until 2020. A fragment of a the object-centric event log is shown in Table~\ref{tab1}. There is one-to-many relationship between order and item and one-to-one relationship between order and invoice. A part of the MVP model of the process is shown in Fig.~\ref{fig-obj-model}. MVPs are graphs in which nodes represent activities and there is an edge between two nodes if there is at least one event in the event log where the source activity is followed by the target activity. In the MVP models, arcs with different colors represent different case notions~\cite{van2019object}. Following the blue and purple arcs, the order and item go through create purchase order, and enter incoming invoice and a few of them leave the process at the \emph{cancel invoice document}.

\begin{table*}[tb]\centering

	\caption{\footnotesize \normalfont\scshape \\A Fragment of an Object-Centric Event Log}\label{tab1}
	\scalebox{1}{%
		\begin{tabular}{|c|c|c|c|c|c|c|}
			\hline
			event  &  \multicolumn{3}{c|}{attributes} & \multicolumn{3}{c|}{object types} \\
			\cline{2-7}
			identifier & activity & timestamp & resource & order & item & invoice\\
			\hline
			0001 &  \emph{create purchase order} & 01-01-2000:00.00 & USER01 & $\left\{o_{1}\right\}$ & $\left\{i_{1}, i_{2}, i_{3}\right\}$  & $\left\{\right\}$\\
			0002 &  \emph{post document} & 01-01-2000:08.05& USER01 & $\left\{\right\}$ & $\left\{\right\}$ & $\left\{inv_{1}\right\}$\\
			... &  ... & ... & ... & ... & ...  & ... \\
			17499 & \emph{enter incoming invoice} & 20-05-2020:15.11 & USER50& $\left\{o_{12000}\right\}$ & $\left\{i_{27005}\right\}$ & $\left\{inv_{800}\right\}$\\
			17500 & \emph{park invoice} & 20-05-2020:15.14 & USER42& $\left\{o_{12000}\right\}$ & $\left\{\right\}$ & $\left\{\right\}$\\
			\hline
	\end{tabular}}
\end{table*}

We also use Object-centric Petri nets in this paper. These discovery techniques focus on solitary processes. There may exist different behaviors in the processes. The variability in this purchasing process motivates us to extend the process cubes notion to support object-centric event logs to compare different processes.

\section{Object-Centric Process Cubes}\label{Process Cubes}
In this section, we formalize the notion of object-centric process cubes. Using the running example, described in Section~\ref{Running Example}, we provide examples for the object-centric process cube notion. A process cube is formed by its structure and an object-centric event log. The structure describes the distribution of the cells and the object-centric event log is used to materialize the cells of cube by events.

\subsection{Object-Centric Event Log}\label{Event Log}

The object-centric event log, shown in Table~\ref{tab1}, represents a collection of events and that is totally ordered based on the timestamp. Each event consists of an event identifier, attributes, and object types (i.e., case notions). To formalize the object-centric event log, we define the universes to be used throughout the paper.

\begin{definition}[Universes]\label{sec:Def1} We define the following universes:
	
	\begin{itemize}
		\item  $\mathbb{U}_{ei}$ is the universe of event identifiers, 
		\item  $\mathbb{U}_{att}$ is the universe of all possible attribute names, 
		\item  $\mathbb{U}_{v}$ is the universe of all possible attribute values, 
		\item  $\mathbb{U}_{vmap}=\mathbb{U}_{att} \nrightarrow \mathbb{U}_{v}$ is the universe of functions mapping attributes on attribute values, 
		\item  $\mathbb{U}_{ot}$ is the universe of all object types, 
		\item $\mathbb{U}_{oi}$ is the universe of object identifiers,
		\item  $\mathbb{U}_{s}=(\mathcal{P}(\mathbb{U}_{v})\cup \mathcal{P}(\mathbb{U}_{oi}))\setminus\left\{\emptyset\right\}$ is the universe of value sets excluding the set containing the empty set,
		\item $\mathbb{U}_{h}=(\mathcal{P}(\mathcal{P}(\mathbb{U}_{v}))\cup \mathcal{P}(\mathcal{P}(\mathbb{U}_{oi})))\setminus\left\{\emptyset,\left\{\emptyset\right\}\right\}$ is the universe of value set collections excluding the set and the set collection containing the empty set,
		\item $type \in  \mathbb{U}_{oi}\rightarrow \mathbb{U}_{ot}$ assigns precisely one type to each object identifier,	
		\item $\mathbb{U}_{omap}$ is the universe of all the object mappings indicating which object identifiers are included per type. $\mathbb{U}_{omap}$ is defined like: 
		\begin{align*}
		\mathbb{U}_{omap} =\{omap\in \mathbb{U}_{ot}\rightarrow \mathcal{P}(\mathbb{U}_{oi})~\mid~\forall _{ot\in \mathbb{U}_{ot}}~ \\
		\forall _{oi\in omap(ot)}~type(oi)=ot\}
		\end{align*}
		\item $\mathbb{U}_{event}=\mathbb{U}_{ei}\times \mathbb{U}_{vmap}\times \mathbb{U}_{omap}$ is the universe of events.
		
	\end{itemize}
\end{definition}

$e=(ei, vmap, omap)\in \mathbb{U}_{event}$ is an event with event identifier $ei$, referring to the objects specified in $omap$, and having attribute values specified by $vmap$. Each row in the Table~\ref{tab1} refers to an event, which contains an event-identifier, attribute values, and object identifiers. Note that $\mathbb{U}_{s}$ does not include $\left\{\emptyset\right\}$ and $\mathbb{U}_{h}$ does not include $\left\{\emptyset,\left\{\emptyset\right\}\right\}$. These values are created after the power set generation. However, it is not meaningful for these sets to contain such values.

\begin{definition}[Object-Centric Event Log]\label{sec:Def2}
	An object-centric event log is a tuple $E=(EI, ATT, OT, \pi_{vmap}, \pi_{omap})
	\in \mathcal{P}(\mathbb{U}_{ei})\times\mathcal{P}(\mathbb{U}_{att})\times\mathcal{P}(\mathbb{U}_{ot})\times\mathcal{P}(\mathbb{U}_{vmap})\times\mathcal{P}(\mathbb{U}_{omap})$ where $\pi_{vmap}\in EI\rightarrow U_{vmap}$ is a function mapping each event to its attribute mapping, and $\pi_{omap}\in EI\rightarrow U_{omap}$ is a function mapping each event to its object mapping. Event identifiers are unique and time is non-descending.
\end{definition}

The first column in the object-centric event log of the running example shows unique event identifiers. Consider $e_1$, the first event in Table~\ref{tab1}. $\pi_{vmap}(e_1)(activity) {=} \text{\emph{create purchase order}}$, $\pi_{vmap}(e_1)(resource) {=} \text{\emph{USER01}}$, $\pi_{omap}(e_1)(order) {=} \left\{o_1\right\}$, and  $\pi_{omap}(e_1)(item) {=} \left\{i_1, i_2, i_3\right\}$.

\subsection{Process Cube Structure}\label{Process Cube Structure}
We define the structure of the process cube independent from the object-centric event log. A process cube structure is fully specified by the set of dimensions.

\begin{definition}[Process Cube Structure]\label{sec:Def4}
	A process cube structure is a triplet $PCS = (D, val, gran)$ where:
	\begin{itemize}
		\item  $D$ is a set of dimensions,
		\item  $val\in D\rightarrow \mathbb{U}_{s}$ is a function associating a dimension to a set of values.
		\item  $gran\in D\rightarrow \mathbb{U}_{h}$ defines a level for each dimension such that for any $d \in D: val(d) = \cup gran(d)$.\\
	\end{itemize}
\end{definition}

A dimension $d$ has a value $val(d)$ and a granularity $gran(d)$. The possible set of values for each dimension is specified by $val(d)$ and a subset of these values exist in a sample of the process cube. For example, $val(item){=}\left\{i_{1}, i_{2}, i_{3}, ..., i_{27005}\right\}$ and $val(activity){=}\left\{\text{\emph{create purchase order}}, ...,\text{\emph{park invoice}}\right\}$. A dimension also has a granularity $gran(d)$ which is a set of sets. For example, $gran(timestamp)$ contains sets such as $T_{2017}$ and $T_{2018}$ each showing all the timestamps in a particular year. These sets form levels based on set inclusion (e.g., $T_{2019}$ dominates $T_{Apr-2019}$). 

The content of the cube is the object-centric event log. Therefore, we make the \emph{process cube structure} and the \emph{object-centric event log} compatible.
\begin{definition}[Compatible]\label{sec:Def5}
	\label{sec:Def5}
	Let $E=(E, ATT, OT, \pi_{vmap}, \pi_{omap})$ be an object-centric event log and $PCS=(D, val, gran)$ be a process cube structure. They are compatible if:
	
	\begin{itemize}
		\item  $D \subseteq OT\cup ATT$, dimensions should correspond to attributes or object types,
		\item  for any $d \in D \cap ATT$ and $e \in E: \pi_{vmap}(e)(d)\in val(d)$,
		\item  for any $d \in D \cap OT$ and $e \in E: \pi_{omap}(e)(d) \subseteq val(d)$.
	\end{itemize}
\end{definition}

In making the process cube structure and the object-centric event log compatible, there is a difference between dimensions that correspond to object types and dimensions that correspond to attributes. This difference arises due to the non-atomic values for object types. Consider the activity as the dimension, which is an attribute, $activity {\in} D {\cap} ATT$, then $\pi_{vmap}(e_1)(activity){\in}\left\{\text{\emph{create purchase order}}, ...,\text{\emph{park invoice}}\right\}$. However, if we consider item as the dimension, which is an object type, $item {\in} D {\cap} OT$, then $\pi_{omap}(e_1)(item) {\subseteq} \left\{i_{1}, ..., i_{27005}\right\}$.

Different operations are possible in process cubes. By applying process cube operations such as slicing, the content of the process cube structure and object-centric event log do not change. We only change the fragment of the cube that we visualize. A process cube view defines the events that are visible for us.

\begin{definition}[Process Cube View]\label{sec:Def5}
	\label{sec:Def5}
	Let $PCS = (D, val, gran)$ be a process cube structure. A process cube view is a pair $PCV = (D_{sel}, sel)$ such that:
	\begin{itemize}
		\item  $D_{sel} \subseteq D$ are the selected dimensions,
		\item  $sel \in D\rightarrow \mathbb{U}_{h}$ is a function selecting the part of the level considered per dimension. The function $sel$ is such that for any $d \in D$: \\
		\begin{itemize}
			\item $sel(d)\subseteq gran(d)$
			\item for any $S_{1}, S_{2} \in sel(d): S_{1}\subseteq S_{2}$ implies $S_{1} = S_{2}$.
		\end{itemize}
	\end{itemize}
	
\end{definition}

\begin{figure}[tb]

	\begin{subfigure}{0.4\textwidth}
		\centering
		% include first image
		\includegraphics[scale=0.13]{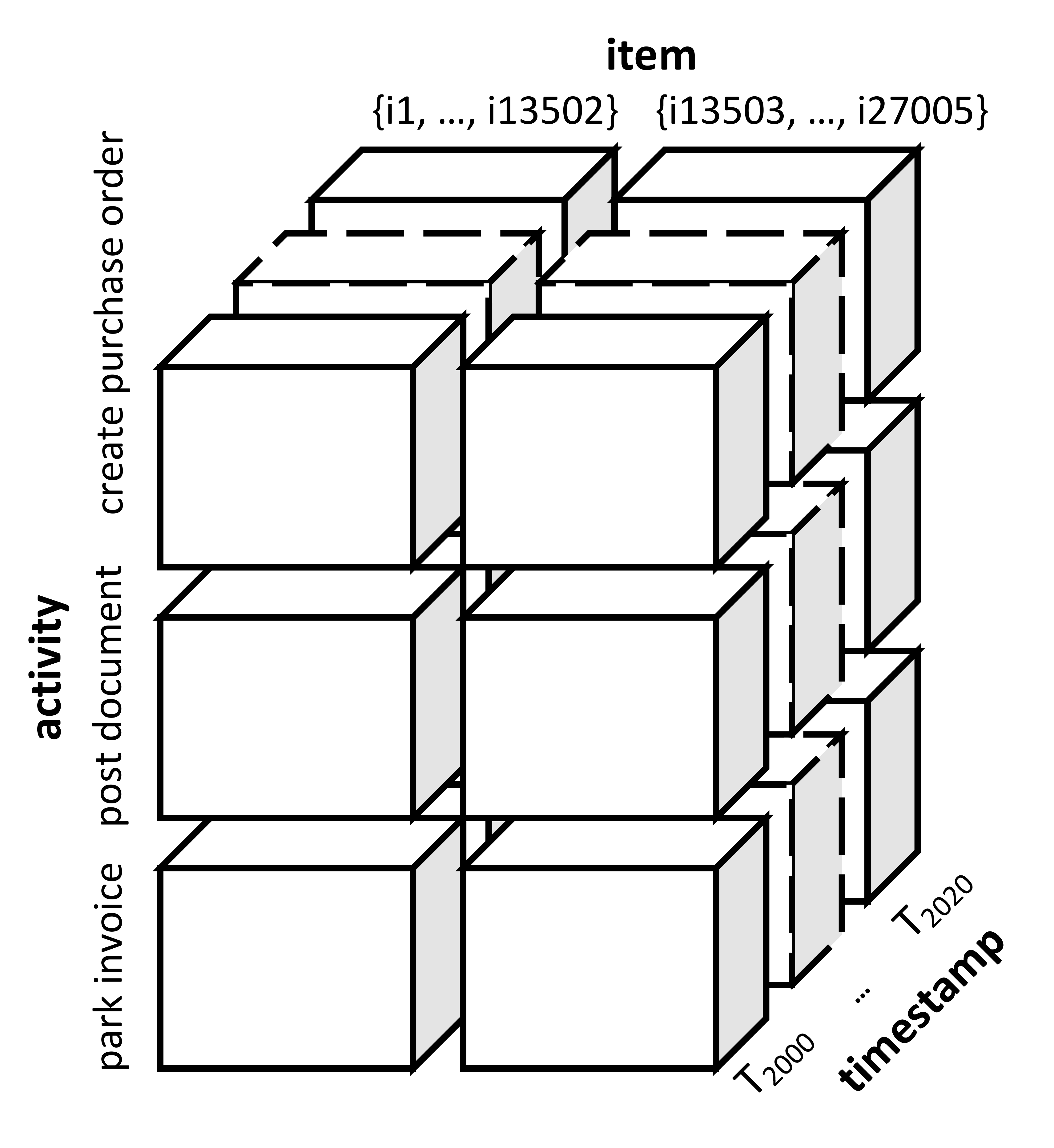}  
		\caption{A process cube view with the granularity of year for the time dimension}
		\label{process-cube-view-year}
	\end{subfigure}
	\begin{subfigure}[t]{0.1\textwidth}
		
	\end{subfigure}
	\begin{subfigure}{0.4\textwidth}
		\centering
		% include second image
		\includegraphics[scale=0.13]{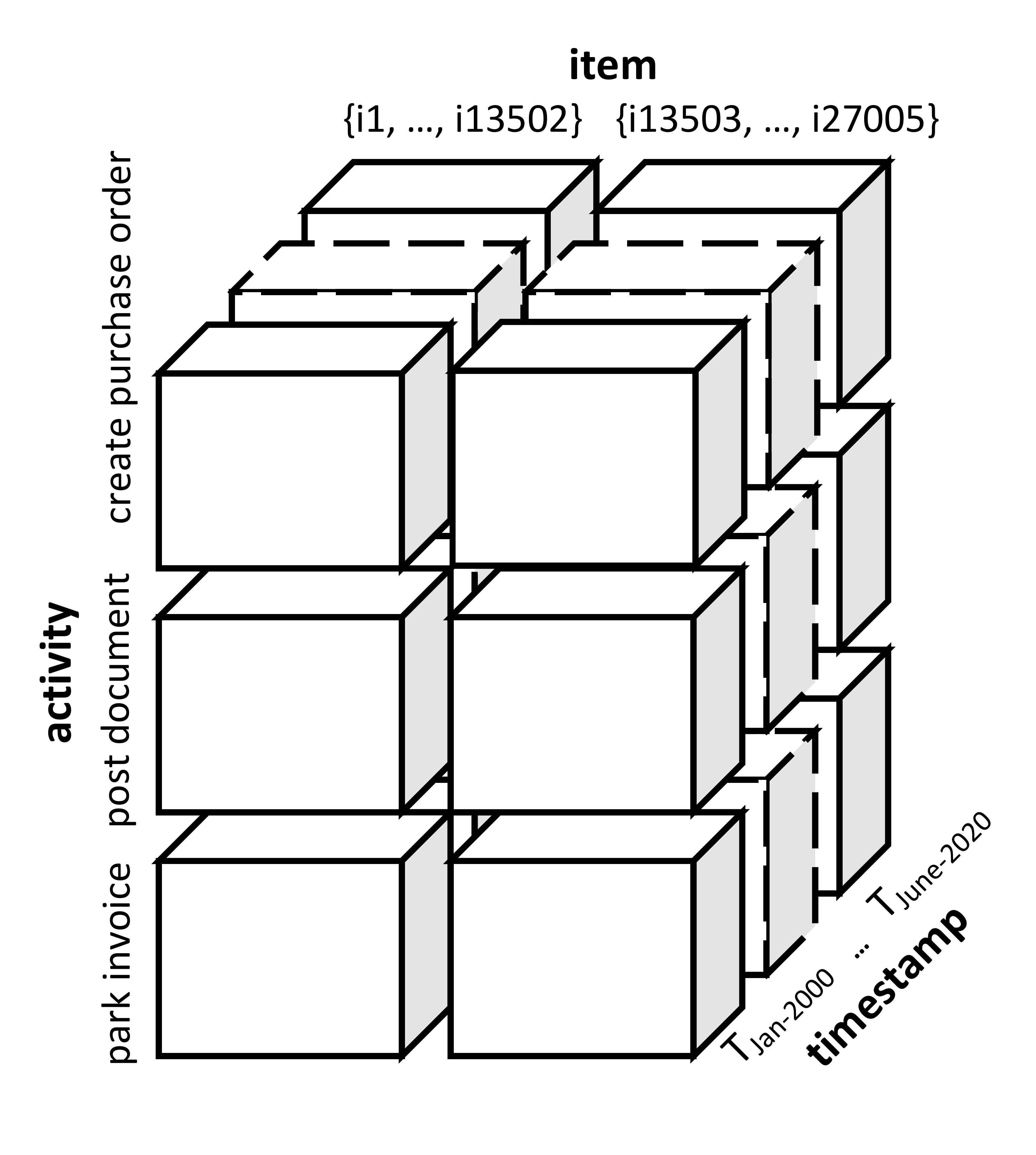}  
		\caption{A process cube view with the granularity of month for the time dimension}
		\label{process-cube-view-month}
	\end{subfigure}
	\caption{Example of different process cube views of the same process cube structure}
	\label{process-cube-views}
\end{figure}

A process cube view is a cube with $|D_{sel}|{\leq} |D|$ dimensions. Fig.~\ref{process-cube-view-year} shows an example of a process cube view with three dimensions. The selected dimensions are activity, item and timestamp ($D_{sel}{=}\left\{item, timestamp, activity\right\}$). Function $sel$ selects values for all dimensions regardless of whether they are in $D_{sel}$ or not. For the $D{\setminus}D_{sel}$ we cannot see the values of $sel$ in the \emph{process cube view}. However, the values of $sel$ exist for these dimensions and these dimensions may have been used in filtering. For example, in slicing the dimension is no longer visible but it is used in filtering. In the process cube view shown in Fig.~\ref{process-cube-view-year}, $sel(timestamp){=}\left\{T_{2000}, ..., T_{2020}\right\}$. Figure~\ref{process-cube-view-month} shows another view where $sel(timestamp){=}\left\{T_{Jan-2000}, ..., T_{June-2020}\right\}$. Through the requirement $S_{1}{\subseteq} S_{2}$ implies $S_{1} {=} S_{2}$, we ensure that elements of $sel(d)$ do not intersect, e.g., $sel(timestamp){=}\left\{T_{Jan-2017}, T_{2017}, T_{2018}, T_{2019}\right\}$ is not possible. By having the process cube view and object-centric event log, we materialize the cells by events. We can extract event logs from cells of the process cube to apply process mining techniques such as process discovery.

\begin{definition}[Materialized Process Cube View]\label{Materialized Process Cube View}
	% \subsubsection{Definition 6 (Materialized Process Cube View).} 
	\label{sec:Def6}
	Let process cube structure $PCS = (D, val, gran)$ and object-centric event log $E = (E, ATT, OT, \pi_{vmap},\pi_{omap})$ be compatible. The materialized process cube for some view $PCV{ =} (D_{sel}, sel)$ of $PCS$ is $M_{E, PCV} = \left\{(c, events(c))~|~c\in cells\right\}$ with 
	$cells = \left\{c\in D_{sel}{\rightarrow} \mathbb{U}_{s}~|~\forall _{d\in D_{sel}} c(d)\in sel(d) \right\}$ being the cells of the cube and 
	
	\begin{align*}
	events(c) =\\
	&\{e {\in} E~|~\forall _{d\in D_{sel} \cap ATT}~ \pi_{vmap}(e)(d) {\in} c(d)\\ 
	&\wedge \forall _{d{\in} D \cap ATT}~ \pi_{vmap}(e)(d) {\in} \cup sel(d)\\
	&\wedge \forall _{d{\in} D_{sel} \cap OT}~ c(d) {\subseteq} \pi_{omap}(e)(d)\\ 	
	&\wedge \forall _{d{\in} D \cap OT}~ \pi_{omap}(e)(d){\subseteq} \cup sel(d)\}\\
	\end{align*}

\end{definition}

In materializing, we add content to the cells. In other words, we create an event log for the cells in the 	process cube view. In a $c\in cells$, each visible dimension is assigned to a value of that dimension, e.g., $c(activity){=}\text{\emph{create purchase order}}$, $c(timestamp){=}T_{2020}$, and $c(item){=}\left\{i_1, i_2, ..., i_{13502}\right\}$. We materialize the cells of the process cube with events. $events(c)$ are all the events corresponding to the first requirement (i.e., $\forall _{d{\in} D_{sel} \cap ATT}~ \pi_{vmap}(e)(d) {\in} c(d)$, $\forall _{d{\in} D_{sel} \cap OT}~ c(d) {\subseteq} \pi_{omap}(e)(d)$) which is different for object types and attributes. The second requirement (i.e., $\forall _{d{\in}D \cap ATT}~ \pi_{vmap}(e)(d) {\in}\cup sel(d)$, $\forall _{d{\in}D \cap OT}~ \pi_{omap}(e)(d){\subseteq}\cup sel(d)$), which is also different for object types and attributes, makes sure that events are not filtered out. Events in the cell can be converted to an object-centric event log and used to apply process mining techniques.

Using earlier formalizations, we define process cube operations such as \emph{slice}.

\begin{figure*}[tb]
	{	\centering
		\includegraphics[scale=0.26]{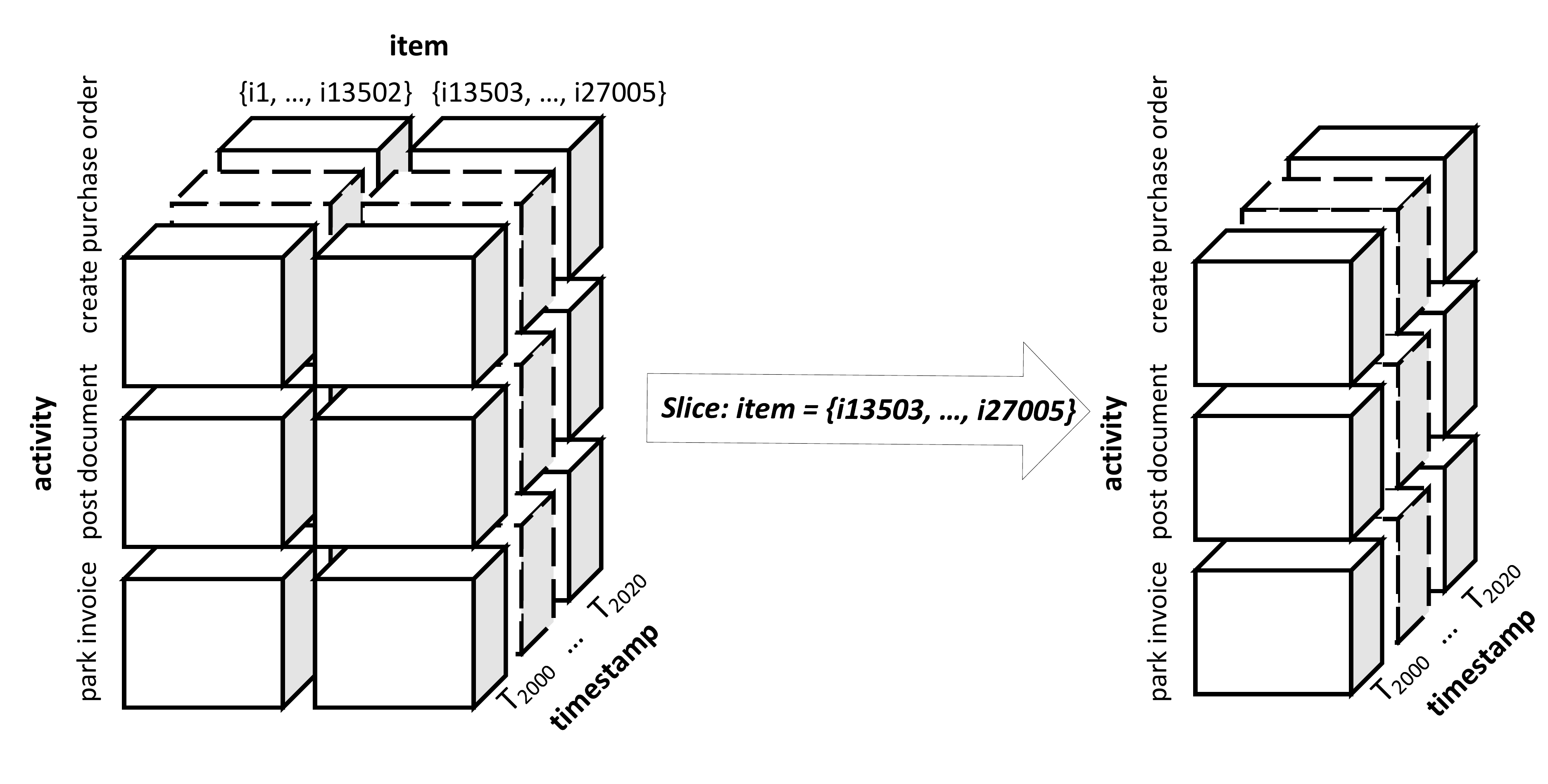}
		\caption{The process cube view after slicing based on dimension item}
		\label{fig-cube-slice}
	}
\end{figure*}
\begin{figure*}[!]
	{	\centering
		\includegraphics[scale=0.26]{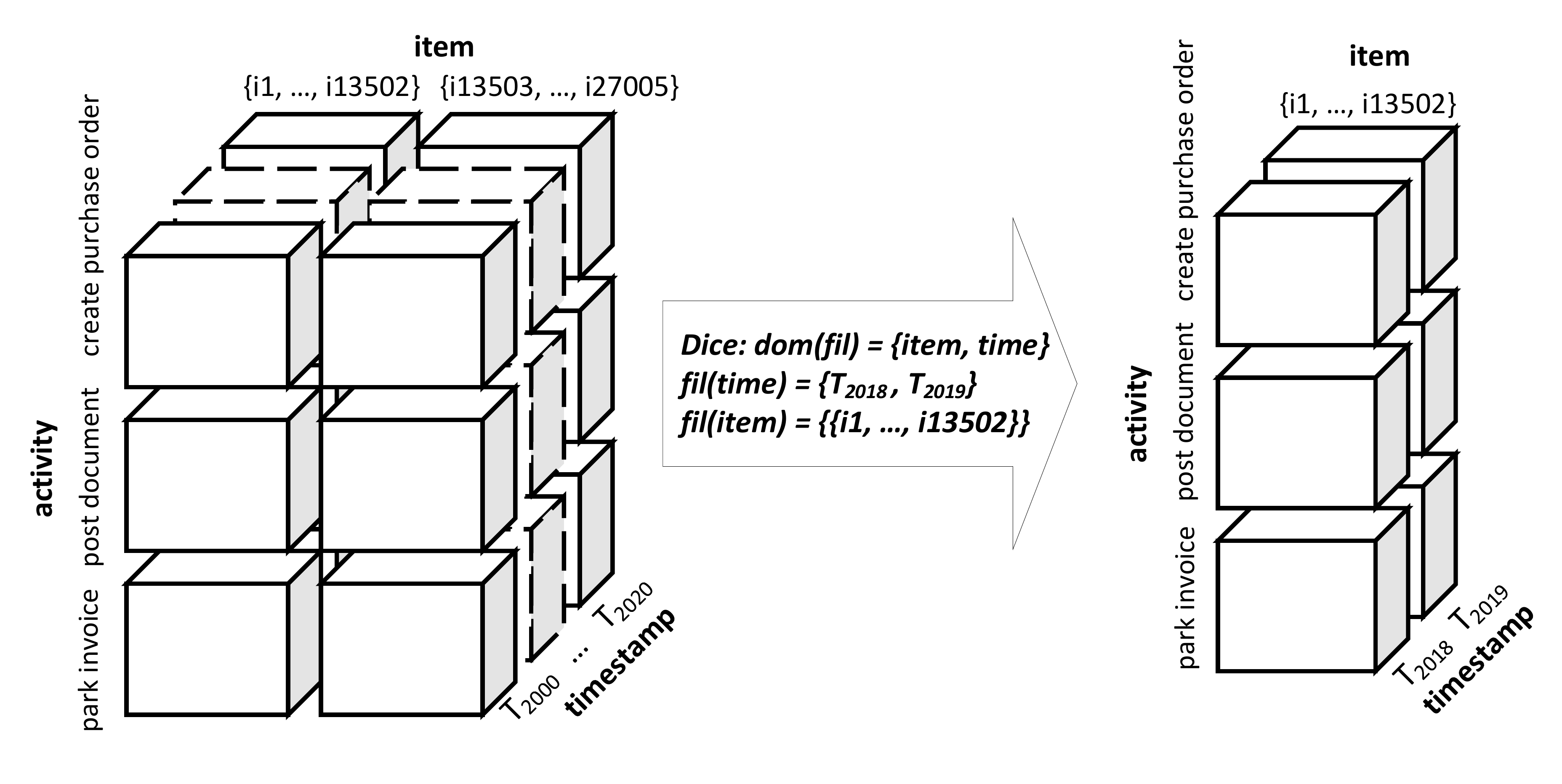}
		\caption{The process cube view after dicing based on dimensions item and time}
		\label{fig-cube-dice}
	}
\end{figure*}

\begin{definition}[Slice]\label{sec:Def7}
	Let $PCS = (D, val, gran)$ be a process cube structure and $PCV = (D_{sel}, sel)$ a view of $PCS$. For any $d \in D_{sel}$ and $V \in sel(d): slice_{d, V}(PCV) = (D_{sel}^{\prime}, sel^{\prime})$ with $D_{sel}^{\prime} = D_{sel}\setminus \left\{d\right\}, sel^{\prime}(d) = \left\{V\right\}$ and $sel^{\prime}(d^{\prime}) = sel(d^{\prime})$ for $d^{\prime} \in D\setminus \left\{d\right\}$.
\end{definition}

Through slicing a new cube view is produced and a dimension \emph{d} is removed from the cube. As shown in Fig.~\ref{fig-cube-slice}, in slicing for the item dimension and value set $\left\{i_{13503}, i_{13504}, ..., i_{27005}\right\}$, the item dimension is removed and only events in which the item is a subset of $\left\{i_{13503}, i_{13504}, ..., i_{27005}\right\}$ remain in the cube view.

\begin{definition}[Dice]\label{sec:Def8}
	Let $PCS = (D, val, gran)$ be a process cube structure and $PCV = (D_{sel}, sel)$ a view of $PCS$. Let $fil\in D_{sel}\nrightarrow U_{h}$ be a filter such that for any $d\in dom(fil): fil(d)\subseteq sel(d)$. $dice_{fil}(PCV)=(D_{sel},sel^{\prime})$
	with $sel^{\prime}(d)=fil(d)$ for $d\in dom(fil)$ and $sel^{\prime}(d)=sel(d)$ for $d\in D\setminus dom(fil)$.
\end{definition}

The difference between dice and slice is that through dicing, no dimension is removed and it limits the values for one or more dimensions. For example dicing is applicable based on two dimensions time and item, as shown in Fig.~\ref{fig-cube-dice}. In this example, we have new process cube view based on two filters: $fil(time)=\left\{T_{2018}, T_{2019}\right\}$ and $fil(item)=\left\{i_{1}, i_{2}, ..., i_{13502}\right\}$.

\begin{definition}[Change Granularity]\label{sec:Def9}
	Let $PCS = (D, val, gran)$ be a process cube structure and $PCV = (D_{sel}, sel)$ a view of $PCS$. Let $d\in D_{sel}$ and $G \in \mathbb{U}_{h}$ such that: $G\subseteq gran(d)$ and $\cup G = \cup sel(d)$. $chgr_{d,G}(PCV)=(D_{sel}, sel^{\prime})$ with $sel^{\prime}(d)=G$, and $sel^{\prime}(d^{\prime})=sel(d^{\prime})$ for $d^{\prime}\in D\setminus\left\{d\right\}$.
	
\end{definition}

The dimensions in process cube view do not change during changing granularity, but the dimensions are shown in a more fine-grained or coarse-grained vision. In Fig.~\ref{process-cube-view-year} the granularity for time dimension is year. However, in Fig.~\ref{process-cube-view-month}, the granularity for time dimension is month. Having different levels of the granularity, we can compare processes in different levels of granularity. 

\section{Implementation}\label{Implementation}
\begin{figure}[!]
	\centering
	\includegraphics[scale=0.25]{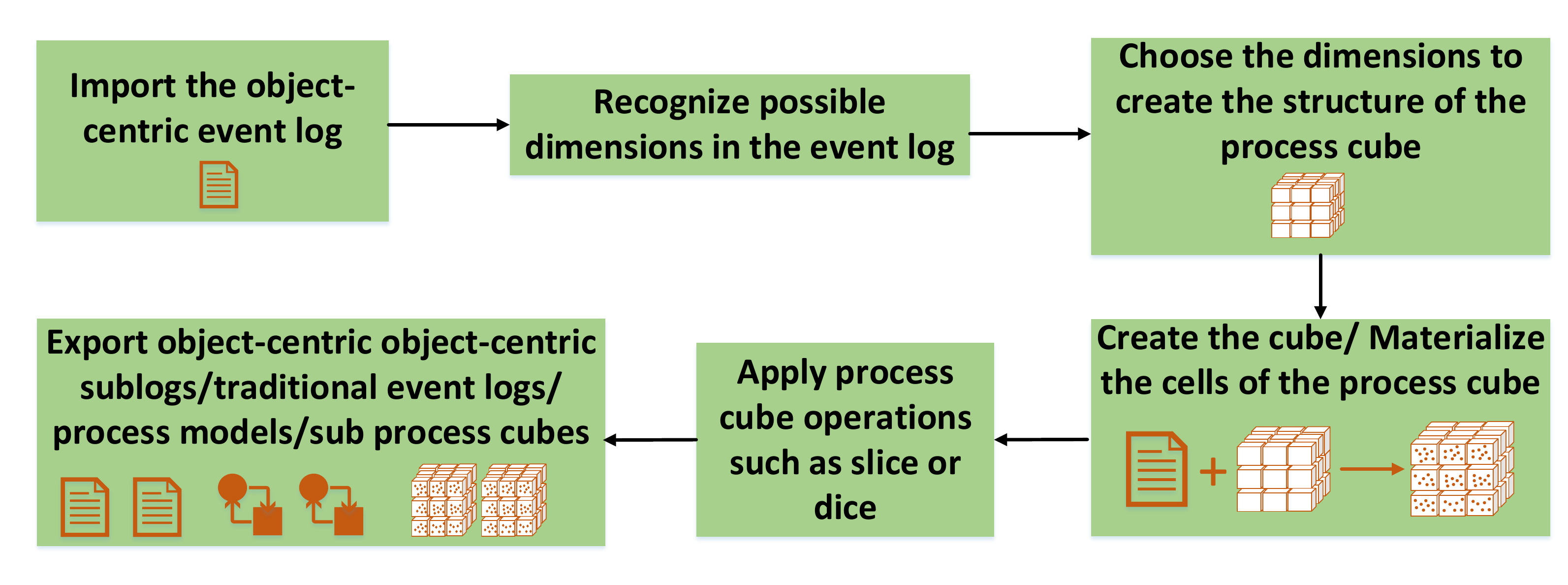}
	\caption{An overview showing how the framework can be used. The input is an object-centric event log. By choosing the dimensions, the user can build the process cube and through process cube operations explore the cube. It is possible to have the output as an event log, a process model, and a sub cube}\label{fig-tool_precedure}
\end{figure}

The approach has been implemented as a standalone Python application \footnote{available in https://gitlab.com/Anahita-Farhang/process-cube} by using PM4PY-MDL. The user can easily install the framework in Python 3.6. The procedure to create a cube in this framework is shown in Fig.~\ref{fig-tool_precedure}. The functionalities of the framework are:

\begin{itemize}

	\item[$\bullet$] It is possible to import object-centric event logs (The framework also accepts CSV and XES formats and automatically converts them to an object centric event log with one case notion).
	\item[$\bullet$] The created cube can be exported in a dump file and stored in the memory for future exploration. 
	\item[$\bullet$] At any point in time, it is possible to export the object-centric event log as an object-centric event log or a traditional event log by selecting a case notion. It is also possible to show the process models of the selected cell(s).

\end{itemize}

\begin{figure*}[!]
	\centering
	\includegraphics[scale=0.5]{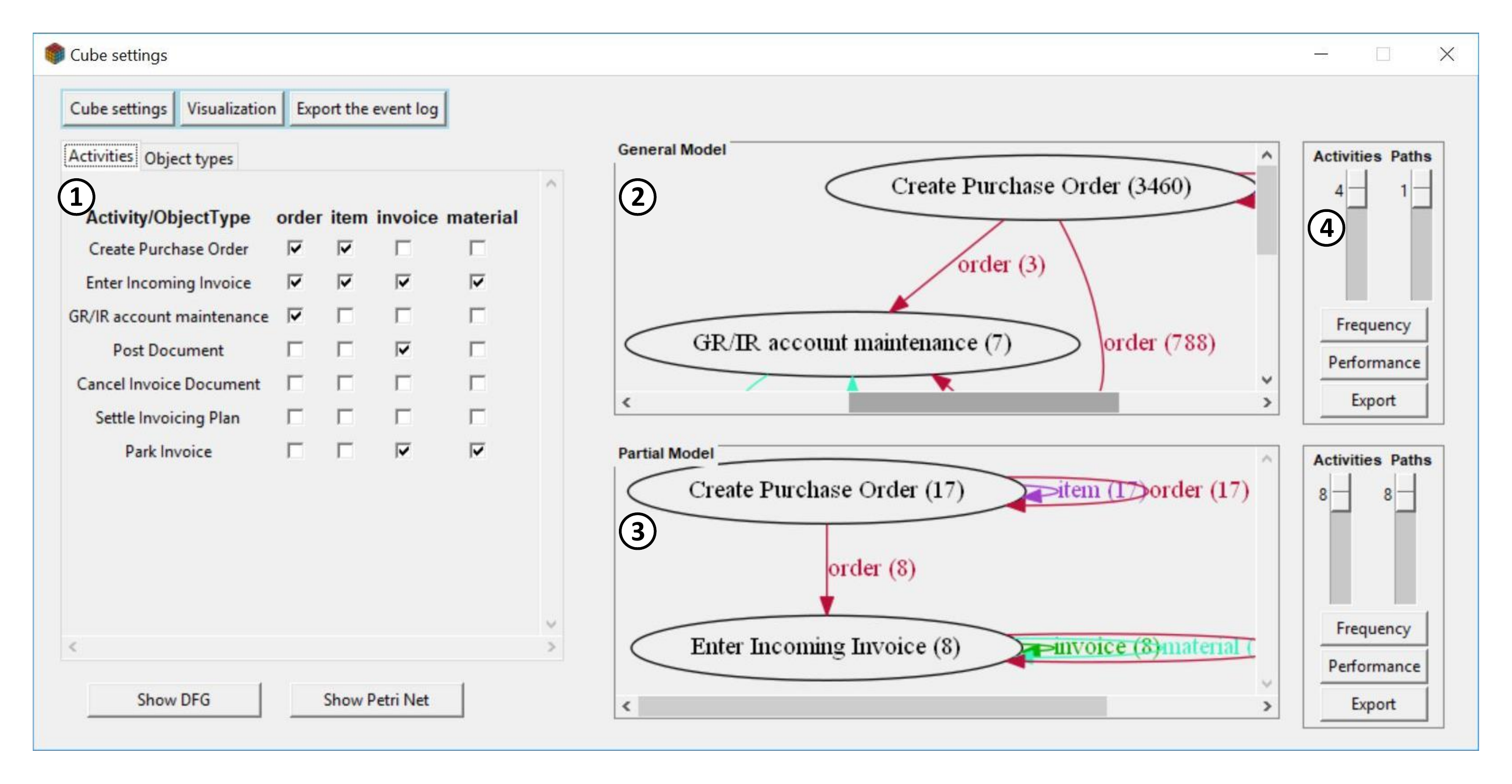}
	\caption{Process Mining Cube: process comparison approach}\label{implementation}
\end{figure*}

It is possible to explore the cube interactively through process cube operations such as slicing. The user can discover an MVP model, enhanced with performance/frequency information, and Object-centric Petri net for each cell. Fig.~\ref{implementation} shows a comparison between MVP models of a dice and the whole cube:

\begin{itemize}
	\item[$\bullet$] In Box~1, it is possible to specify for each object type the activities that are considered for that object type. For example, for the activity \emph{create purchase order}, \emph{orders} are involved, but \emph{items} are not involved in the \emph{post document}.
	\item[$\bullet$] The MVP model of the whole cube is shown in Box~2 by considering the filtering options in Box~1. MVP models are DFGs with colored arcs. In this figure, the color of the arcs related to the \emph{order} is red.
	\item[$\bullet$] The MVP model of the specific \emph{slice}/\emph{dice} of the cube is represented in Box~3. Putting this model near the whole cube's model makes the comparison easier.
	\item[$\bullet$] In Box~4, the user can change the frequency of the nodes and edges appearing in the MVP model. There is a performance annotated version of MVP models that is reachable only by clicking on performance. It is possible to export the object-centric process model as an image in the desired address.
	
\end{itemize}
\section{Evaluation}\label{Evaluation}

In this section, we analyze the scalability of the object-centric process cube tool. To assess the performance of our approach, we measure the scalability of the approach from two perspectives: creation time (e.g., the time required to create and materialize cells of the cube), and the loading time (e.g., the time required to import the cube). Results, shown in Fig.~\ref{performanceanalysiscreatingthecube}, have been done in three different settings: the time for creating/loading the cube in terms of the number of events in the event log (while keeping the number of object types and attributes constant), the number of object types in the event log (while keeping the number of events and attributes constant), and the number of attributes in the event log (while keeping the number of events and object types constant). Performance analysis of the cube with different settings shows the time required for creating/loading the cube increases linearly, non-linearly, and linearly when increasing number of events, object types, and attributes respectively.
The proposed framework creates a process cube for an event log with 17500 events in almost 4 minutes. 
\pgfplotstableread[col sep=space,row sep=newline,header=true]{
	x   y
	1500  12
	3500  36
	5500  60
	7500  81
	9500  111
	11500 147
	13500 189
	15500 227
	17500 271 
}\createthecubeevents

\pgfplotstableread[col sep=space,row sep=newline,header=true]{
	x   y
	1 4
	2 14
	3 75
	4 271
}\createthecubeobjecttypes

\pgfplotstableread[col sep=space,row sep=newline,header=true]{
	x   y
	1 80
	2 150
	3 208
	4 271
}\createthecubeattributes

\pgfplotstableread[col sep=space,row sep=newline,header=true]{
	x   y
	1500  174
	3500  420
	5500  621
	7500  910
	9500  1086
	11500 1314
	13500 1588
	15500 1854
	17500 2005 
}\loadthecubeevents

\pgfplotstableread[col sep=space,row sep=newline,header=true]{
	x   y
	1 195
	2 480
	3 1001
	4 2005
}\loadthecubeobjecttypes

\pgfplotstableread[col sep=space,row sep=newline,header=true]{
	x   y
	1 587
	2 1090
	3 1560
	4 2005
}\loadthecubeattributes

\begin{figure*}[!]
	\begin{minipage}{0.45\textwidth}
		\resizebox{0.8\columnwidth}{!}{%
			\begin{tikzpicture}
			\begin{axis}[xlabel={Number of events}, ylabel={Create the cube (sec)}]
			\addplot[smooth] table {\createthecubeevents};
			\addplot[only marks,mark=*,mark options={color=purple}] table {\createthecubeevents};
			\end{axis}
			\end{tikzpicture}
		} \\
		a.1) The difference in time required for creating the cube for different numbers of events: n\_object types = 4, and n\_attributes = 4\\
	\end{minipage}
	\begin{minipage}[t]{0.1\textwidth}
		~
	\end{minipage}
	\begin{minipage}{0.45\textwidth}
		\resizebox{0.8\columnwidth}{!}{%
			\begin{tikzpicture}
			\begin{axis}[xlabel={Number of events}, ylabel={Load the cube (msec)}]
			\addplot[smooth] table {\loadthecubeevents};
			\addplot[only marks,mark=*,mark options={color=blue}] table {\loadthecubeevents};
			\end{axis}
			\end{tikzpicture}
		} \\
		a.2) The difference in time required for loading the cube for different numbers of events: n\_object types = 4, and n\_attributes = 4\\
	\end{minipage}
	\begin{minipage}{0.45\textwidth}
		\resizebox{0.8\columnwidth}{!}{%
			\begin{tikzpicture}
			\begin{axis}[xlabel={Number of object types}, ylabel={Create the cube (sec)}]
			\addplot[smooth] table {\createthecubeobjecttypes};
			\addplot[only marks,mark=*,mark options={color=purple}] table {\createthecubeobjecttypes};
			\end{axis}
			\end{tikzpicture}
		}\\
		b.1) The difference in time required for creating the cube for different numbers of object types: n\_events = 17500, and n\_attributes = 4\\
	\end{minipage}
	\begin{minipage}[t]{0.1\textwidth}
		~
	\end{minipage}
	\begin{minipage}{0.45\textwidth}
		\resizebox{0.8\columnwidth}{!}{%
			\begin{tikzpicture}
			\begin{axis}[xlabel={Number of object types}, ylabel={Load the cube (msec)}]
			\addplot[smooth] table {\loadthecubeobjecttypes};
			\addplot[only marks,mark=*,mark options={color=blue}] table {\loadthecubeobjecttypes};
			\end{axis}
			\end{tikzpicture}
		}\\
		b.2) The difference in time required for loading the cube for different numbers of object types: n\_events = 17500, and n\_attributes = 4\\
	\end{minipage}
	\begin{minipage}{0.45\textwidth}
		\resizebox{0.8\columnwidth}{!}{%
			\begin{tikzpicture}
			\begin{axis}[xlabel={Number of attributes}, ylabel={Create the cube (sec)}]
			\addplot[smooth] table {\createthecubeattributes};
			\addplot[only marks,mark=*,mark options={color=purple}] table {\createthecubeattributes};
			\end{axis}
			\end{tikzpicture}
		}\\
		c.1) The difference in time required for creating the cube for different numbers of attributes:  n\_events= 17500 , and  n\_object  types = 4.\\
	\end{minipage}
	\begin{minipage}[t]{0.1\textwidth}
		~
	\end{minipage}
	\begin{minipage}{0.45\textwidth}
		\resizebox{0.8\columnwidth}{!}{%
			\begin{tikzpicture}
			\begin{axis}[xlabel={Number of attributes}, ylabel={Load the cube (msec)}]
			\addplot[smooth] table {\loadthecubeattributes};
			\addplot[only marks,mark=*,mark options={color=blue}] table {\loadthecubeattributes};
			\end{axis}
			\end{tikzpicture}
		}\\
		c.2) The difference in time required for loading the cube for different numbers of attributes: n\_events = 17500, and n\_object types = 4.\\
	\end{minipage}
	\caption{Performance analysis of creating (diagrams with red dots)/loading (diagrams with blue dots) the cube for the proposed approach based on a) the number of events b) the number of object types, and c) the number of attributes}
	\label{performanceanalysiscreatingthecube}
\end{figure*}

\section{Conclusion}
\label{Conclusion}
This paper bridges the gap between object-centric process mining and process comparison. Therefore, we proposed an object-centric process cube that organizes data through dimensions referring to case notions (i.e., process instances) and attributes in the event log. The proposed framework allows the users to explore the object-centric event logs interactively through the process cube operations such as slice. Furthermore, the proposed framework, which is publicly available, is able to discover object-centric process models from object-centric event logs extracted from the cells of the cube, which speeds up process comparison. For the future work, we aim to add more features related to performance to the object-centric process cube framework to facilitate the process analysis
  \newpage
\section*{Acknowledgments}
\label{Acknowledgments}
We thank the Alexander von Humboldt (AvH) Stiftung for supporting our research. Funded by the Deutsche Forschungsgemeinschaft (DFG, German Research Foundation) under Germany's Excellence Strategy–EXC-2023 Internet of Production – 390621612.

\bibliographystyle{IEEEtran}
\bibliography{biblio}
%\begin{biography}[{\includegraphics[width=1in,height=1.25in,clip,keepaspectratio]{mshell}}]{Michael Shell}


%\vfill


\end{document}